\documentstyle[11pt,aaspp4]{article}

\lefthead{Natalucci et al.}
\righthead{A new bursting X-ray transient: SAX~J1750.8-2900}

\begin{document}

\title{A new bursting X-ray transient: SAX~J1750.8-2900}


\author{L.Natalucci$^{1}$, R.Cornelisse$^{2,3}$, A.Bazzano$^{1}$,
M.Cocchi$^{1}$,       P. Ubertini$^{1}$, J.Heise$^{2}$, \\ J.J.M.
in~'t~Zand$^{2}$        and E.Kuulkers$^{2,3}$}

\affil{ {  $^{1}$ -- Istituto di Astrofisica Spaziale del CNR, via
       del Fosso del Cavaliere, 00133 Roma, Italy} \\        
       {$^{2}$ -- Space Research Organization Netherlands, Sorbonnelaan 2,
       3584~CA Utrecht, The Netherlands}  \\
       {$^{3}$ -- Astronomical Institute, Utrecht University, P.O. Box 80000, 
       3508 TA Utrecht, The Netherlands} 
}

\altaffiltext{1}{e-mail address: lorenzo@ias.rm.cnr.it}


\begin{abstract} 
We have analysed in detail the discovery measurements of the X-ray burster
SAX~J1750.8-2900 by the Wide Field Cameras on board {\em BeppoSAX}
in spring 1997, at a
position $\sim$1.2 degrees off the Galactic Centre. The source was in
outburst on March 13th when the first observation started and showed 
X-ray emission for $\sim$~2 weeks. A total of 9 bursts were detected,
with peak intensities varying from $\approx$0.4 to 1.0 Crab in the 2-10 keV
range. Most bursts showed a fast rise time ($\approx$1~s), an exponential decay
profile with e-folding time of $\approx$5~s, spectral softening during 
decay, and a spectrum which is consistent with few keV blackbody radiation.
These features identify them as type-I X-ray bursts of
thermonuclear origin.  The presence of type~I
bursts and the source position close to the Galactic Centre favours the
classification of this object as a neutron star low mass X-ray binary. X-ray
emission from SAX~J1750.8-2900 was not detected in the previous and subsequent
Galactic bulge monitoring, and the source was never seen bursting again. 

\end{abstract}



\section{Introduction}
A long term program to survey the 40$\times$40 degrees around the Galactic 
Centre started on mid
1996 with the large field of view instruments on board the {\em BeppoSAX}
satellite (Wide Field Cameras, hereafter WFC).  Previous surveys of
the region with similar instruments were limited by the lack of the
combination of sufficiently long, repeated exposures and the wide angular
coverage. 
In the last 10 years, however, the use of the coded mask imaging technique
increased the total number of known X-ray emitters in the region (\cite{Ski93};
\cite{Var97}) stimulating detailed measurements of individual sources and in
turn their identification at different wavelengths. 

The Galactic Bulge monitoring program carried out by {\em BeppoSAX}~WFC
in the energy range 2-30 keV
has been especially prolific in the study of X-ray burst sources,
increasing substantially (by about 50\% in 2.5 years) the number of objects of
this type which were known originally in this region. As of January 1999 it 
led to the discovery of 6 new burst sources and, in addition, found burst
emission from 7 already known sources (\cite{Hei99}; \cite{Ube99a};
\cite{Coc98a} for earlier results) in a total time exposure of 
$\approx$2.5$\times$10$^{6}$s.
The new transient sources show dim X-ray outburst
episodes during $\sim$1 to a few weeks, with peak fluxes generally below a few
10$^{37}$~erg/s at 10~kpc distance (\cite{Hei99}). From one of these sources,
SAX~J1808.4-3658 (\cite{Int98}) a modulation period of 2.5 ms was discovered
by {\em RXTE} during a second outburst (\cite{Wij98}, \cite{Cha98}).   

Here we report results of one of these previously unknown transients showing
bursting behaviour, discovered by the WFC on March 18th, 1997 
(\cite{Baz97a}, \cite{Hei97}) in a celestial position
1.2 degrees  off the Galactic Center. In particular, we analyse the spectral and 
temporal behaviour of the persistent emission and characterise the burst
emission properties to determine the nature of the transient.


\section{Observations and data analysis}

The  Wide Field Cameras experiment on board the {\em BeppoSAX}
satellite  comprises 2 identical coded aperture multi-wire
proportional counter detectors viewing opposite sky directions
(\cite{Jag97}), each one featuring a field of view of 40$\times$40 
degrees full width to zero
response (i.e., 3.7$\%$ of the sky) and an angular resolution of 5 arcmin. The
source location accuracy depends on the signal-to noise ratio and is
0.7 arcmin at best (99$\%$ confidence level). The energy range
is 2-30~keV on-axis and the time resolution is 0.5~ms. The field of view (FOV)
is the largest of any flown X-ray imaging device with 
arcmin resolution, which allows
for the search of short duration and/or weak transient events. 
The on-axis sensitivity for the Galactic Bulge field is $\approx$10~mCrab in
10$^{4}$s observing time. 
Detector data contain a superposition of background and of
multiple source shadowgrams, the latter resulting from the coding of the sky
object image with the instrument aperture pattern. The reconstruction of the
sky image for point-like sources involves an algorithm that consist of a cross
correlation of the detector data with the aperture (see e.g. \cite{Car87}). 
The position and intensity of any point source is determined by folding a sky
model distribution through a point spread function (PSF), using iterative
$\chi$~$^{2}$ minimisation (\cite{Jag97}). For WFC this can be carried out in
each individual energy channel. The
full-width at half maximum of the PSF is smallest on axis at $\approx$5~arcmin.
SAX~J1750.8-2900 is located only 1.2 degrees off the Galactic Centre and so in
the most sensitive region for this type of observations. The Galactic Bulge
was observed during spring 1997 for 5~$\times$10$^{5}$~s, spread out along 28
days. 

Burst phenomena are systematically searched in data from both 
cameras using time profiles of the total detector over the entire energy
range with a time resolution of 1~s. When a burst occurs 
a reconstructed sky image is generated for the burst duration and different
sky images corresponding to longer time exposure are generated  for
intervals just before and after burst. This allows to resolve the point
source responsible for the intensity increase revealed in detector ratemeters.
In crowded fields and in some not evident case an image subtraction is
necessary to facilitate identification of bursting sources in the FOV.


\section{Transient source position and lightcurve}

\newcommand{\SAX}{{SAX~J1750.8-2900}}

Figure 1 shows the error region for  SAX~J1750.8-2900.
The best fit position is R.A.~=~17h~50m~24s, Dec~=~-29$^{\circ}$~02'~18"
(equinox 2000.0), with an error radius of 1 arcmin (99\% confidence).
This is a position refined from the previously published value which resulted 
from a quick-look analysis (\cite
{Hei97}). The deviation between both values is 0.4 arcmin. Also shown are the
positions of two X-ray bursts that were observed simultaneously with the active
phase of SAX~J1750.8-2900, showing that they result from a position
coincident with the transient.  In March 1992 the {\em ROSAT}~PSPC observed
the region around \SAX ~four times (between MJD~48685.09 and MJD~48691.63 with
exposure times up to 1976~s), during a raster scan
of the Galactic Centre region. No source was detected
during these observations within the 99\% confidence error
box of \SAX. The source 1RXP~J175029-2859.9 lies 1.5 arcmin outside the 
\SAX~error box. This close-by source was marginally detected at
$6.2\pm1.9$~$\times10^{-3}$ counts~s$^{-1}$ in the second observation 
(between MJD~48685.36
and MJD~48685.39). From this result we can derive an upper limit of 
$\sim$~3~$\times10^{-12}$ erg~cm$^{-2}$~s$^{-1}$ on the
soft X-ray emission (0.5-2.0~keV) of \SAX~during quiescence. 

\begin{figure}
\plotone{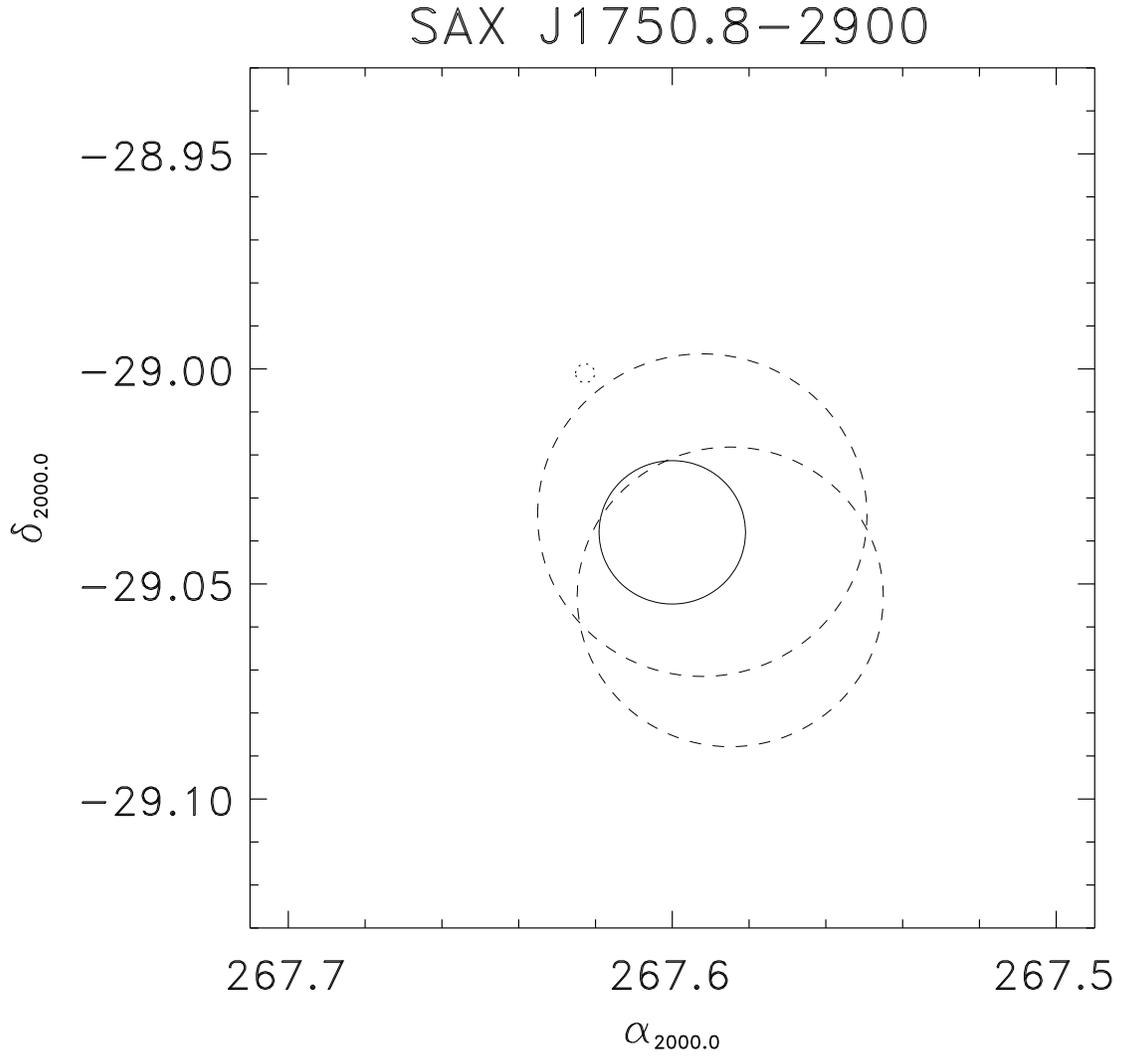}
\caption{
      Position of SAX J1750.8-2900 computed by analysis
      of both persistent and burst images. Dashed contours
      are the source positions estimated from two X-ray bursts, and
      the solid line circle (1 arcmin radius) is the error circle of the
      transient source. Also shown (dotted circle) is the position of the  
      ROSAT source 1RXP~J175029-2859.9. 
      All contours represent 99\% confidence. 
   \label{fig1}}
\end{figure}

\placefigure{fig1}

The {\em RossiXTE} All Sky Monitor (\cite{Lev96}) data retrieved    
from a public dataset 
(provided by the {\em RXTE}/ASM team, http://space.mit.edu/XTE)
shows the onset of a fast rise,
exponential decay outburst of SAX~J1750.8-2900 peaking at 
$\sim$~120$\pm$40~mCrab in the 2-10 keV range, 
starting close to MJD~50518 (two days before the initial WFC
observation) and lasting $\approx$~2 weeks. This transient behaviour is
supported by WFC later detections as previously reported by \cite{Int97}.

\begin{figure}
\plotone{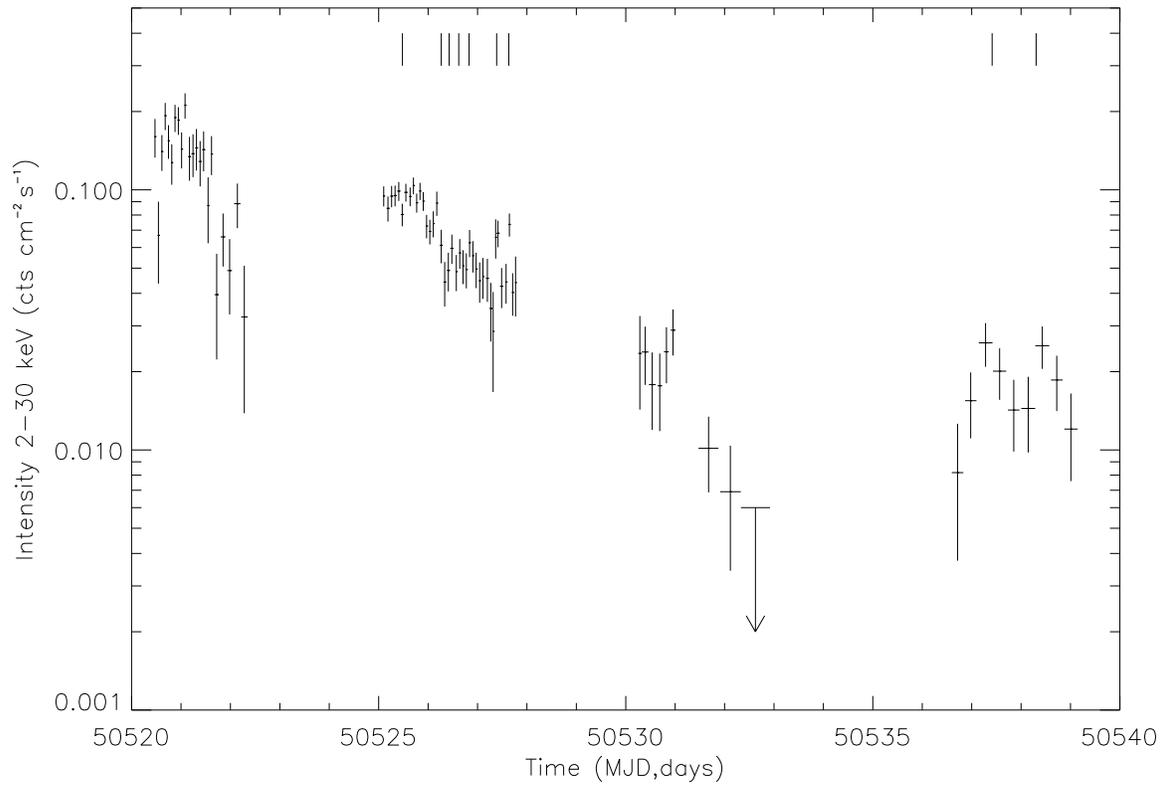}
\caption{
      Light curve of the persistent emission from SAX~J1750.8-2900 in
      the 2-30 keV band.  The markers indicate the epoch of the observed
      bursts.  
    \label{fig2}}
\end{figure}

\placefigure{fig2}

The X-ray persistent emission of SAX~J1750.8-2900
(i.e., the emission detected during time intervals excluding bursts) was
measured by the WFC starting from March~13th, 1997. The source flux was
initially at a level of $\approx$~70 mCrab in the energy band 2-30 keV. The 
light curve detected in the 2-30 keV range is shown in Fig.~2, along with the
burst occurrence time. The outburst profile decay is close to 
exponential, with intensity changes on the time scale of hours.  
The first two days measurements (when the source was more
luminous) are affected by relatively large errors caused by the off-axis 
position of the source. The average luminosity in the
2-30 keV band, calculated for 10 kpc distance is
$\approx$3$\times$10$^{37}$ erg~s$^{-1}$ on March 13th. The flux was 
about $\sim$~1.5 times weaker five days later, when the source
was seen bursting for the first time, and dropped to less than 
$\sim$~3~mCrab
($\approx$~10$^{36}$ erg~s$^{-1}$ at 10 kpc) on March~25th.
The source was again visible at $\sim$~10~mCrab on March 30-31, when 
two more bursts were detected.

We fitted the emission spectra detected in the 2-30 keV band during six
observing periods between MJD~50520 and MJD~50531
using a few spectral models.  The results obtained for power law and thermal
bremssthralung (both with low energy absorption) are shown in Table~1.
The spectra can be described either by a power law shape having a photon
index $\Gamma$~$\approx$~2.5 and extinction parameter
${N}_{H}$~$\approx$~6~$\times$10$^{22}$~cm$^{-2}$, or by bremssthralung
emission with kT~in the 3 to 10~keV range and 
${N}_{H}$~$\approx$~2.5~$\times$10$^{22}$~cm$^{-2}$. The spectra cannot be
fitted satisfactorily with single component blackbody emission.

The fit results give indication that the source has experienced
spectral softening during the outburst decay. By performing an F-test
on the two spectra taken at MJD~50520 and MJD~50527 we find that the 
probability that there is no softening is less than 1\%.
If the X-ray emission mechanism is thermal (as observed in many X-ray 
bursters) the softening could be ascribed to a temperature variation 
of the electron plasma, possibly due to the decrease in the accretion flow.

\placetable{tbl1}


\section{The X-ray bursts}

A total of 9 X-ray bursts were detected from SAX~J1750.8-2900 during an 
overall time span of 14 days in spring 1997. The first one (the faintest
observed) occurred on MJD~50525.48150,  with a peak flux of $\approx$0.4
Crab.  7 out of 9 bursts were detected during three days from March 18th
(see Table 2 for burst occurrence times), having 
similar bolometric fluences in the range $\approx$2 to 3$\times$10$^{-7}$
erg~cm$^{-2}$. In Fig.3 burst profiles in two energy bands are plotted
for two of these events. 

The study of the burst frequency is limited by the fact that 
during observation the effective exposure time is only a fraction 
($\approx$60$\%$) of the total pointing time, due to earth occultations and 
other shorter non-coverage periods. The observed values of time
intervals are in fact an upper limit to the real burst interval time. It is
then possible that SAX~J1750.8-2900 made bursts during the 
first observation period when its persistent flux was above $\sim$50 mCrab.
In spite of this, there is evidence that the burst
frequency decreased when the source flux dropped below $\sim$20 mCrab, i.e. 
in observations performed after MJD 50530 (see Fig.2).

The primary question concerning the bursts is whether they 
are type I X-ray bursts. 
All burst profiles detected from SAX~J1750.8-2900 show a
fast rise ($\approx$~1~s), exponential decay shape, with e-folding time in the 
range $\approx$~5-10~s (see Fig.3). 
The decay times in the energy band 8-26~keV are systematically shorter
than those observed in the band 2-8~keV. However, the spectral softening 
cannot be proven by examining the individual bursts, due to 
the large statistical error (see Table 2). 
In order to increase significance we summed up the profiles of
the last 7 bursts in the energy bands 2-8 and 8-26 keV, with a time
resolution of 0.1~s. The first two bursts were excluded because their 
detection was affected by the earth atmosphere.
The start channel of each burst was determined as the first point in the 
time profile which differed more 
than $\sim$4$\sigma$ from the mean persistent emission. The fit of the 
two profiles obtained with an exponential function gives an e-folding 
decay time $\tau$~=~5.3$\pm$0.7~s in the low energy band and
$\tau$~=~2.8$\pm$0.2~s in the high energy band, and proves that spectral
softening is occurring during burst decay. Together with a consistency of the
burst spectrum with that of a few keV blackbody emission (see Table 2) this 
identifies the bursts as type-I.

Among the bursts detected, there is no clear evidence of an X-ray burst with
double peaked or flat profile, which might have suggested saturation of the 
luminosity to near-Eddington level and resulting photospheric radius expansion
(\cite {Lew95}). However we can derive an upper limit on the source distance
assuming that the maximum burst luminosity was below Eddington. The maximum
observed peak flux (burst F, see Table 2) is consistent with a 3~$\sigma$
upper limit of  $\approx$~7~kpc. 

Burst spectra are rather soft and generally compatible with blackbody
emission having colour temperatures between 2 and 3~keV (see Table 2).
Under given assumptions (\cite {Lew93}) the effective temperature
${T}_{eff}$ and the bolometric flux of a burst can determine the ratio between
the blackbody radius ${R}_{bb}$ (that is, the radius of the emitting sphere)
and the distance d of the neutron star. Assuming d=10~kpc and the observed 
colour temperatures as ${T}_{eff}$, and not correcting for gravitational
redshift the measured blackbody radius is $\approx$~8~km. 
For the above upper limit of 7 kpc, this value of ${R}_{bb}$ scales to a
corresponding upper limit of $\approx$~6~km. This value could be underestimated,
due to the uncertainties in the relationship between colour and effective 
temperature. If,
as suggested by \cite{Ebi87} the colour temperature exceeds ${T}_{eff}$ by a
factor $\approx$1.5, then the neutron star radius should be at least two
times the  measured blackbody radius. These values are therefore consistent
with a neutron star nature of the compact object.

\placetable{tbl2}


\section{Discussion}

\subsection{Burst emission properties}

In the simplest interpretation of the thermonuclear flash model which
successfully explains type-I X-ray bursts (\cite {Lew95} for review)
the matter accreted onto a neutron star surface prior to an observed 
type-I burst is converted into nuclear fuel and the fraction of the total
accreted energy available for burning depends on the actual reaction 
process and fuel composition. If the thermonuclear flash is isotropic and
the accreted material is totally converted into fuel, the ratio
between the mass and radius of the NS is given by
${M}_{*}$/${R}_{10km}$~=~(0.01-0.04)*$\alpha$, where 0.01 and 0.04
hold for helium and hydrogen burning respectively. Here ${M}_{*}$ is the
mass of the compact object in units of solar masses, ${R}_{10km}$ is the NS
radius in units of 10~km and $\alpha$ is the ratio between the
bolometric flux of the persistent emission (integrated over the burst interval)
and the bolometric fluence of the burst. A 1.4~${M}_{\odot}$ neutron star would
then result in a value of  $\alpha$~$\approx$~100 for pure helium burning. For
the four shortest observed burst times intervals B-C,C-D,D-E and F-G (see
Table~2) we estimated the $\alpha$ parameter 
and found values of 85$\pm$20, 120$\pm$30, 170$\pm$30 and 210$\pm$40
respectively. These intervals are monotonically   
increasing from 4.1 to 5.9~h on a time scale of 1.5 days, and all the
related bursts show similar profiles and fluences. This suggests that 
perhaps no bursts were missed in between. The fast
rise time of the bursts ($<$~2~s) and the measured values of $\alpha$ seem to
favour a pure helium flash respect to combined hydrogen-helium shell burning.
(\cite {Lew93}). 

\subsection{SAX~J1750.8-2900 and the transients of the Galactic Bulge}

Most X-ray burst sources known so far are type-I bursters associated with
low-mass X-ray binaries (LMXBs) containing old, weakly magnetized neutron
stars, and show concentration in the direction of the Galactic Centre (\cite
{Van95} ). They can be persistent (though variable) or transient, and may have
recurrence periods with nearly constant burst activity, like the recently
studied GS~1826-338 (\cite {Ube99b}), or conversely show only episodic  burst
emission, like for example SLX~1735-26 (\cite {Baz97b}) and XTE~J1709-267
(\cite {Coc98b}). 
Correlation between burst frequency and persistent emission is not an
uncommon feature. In general, type-I bursts are observed when the source
persistent luminosity is comprised between $\sim$10$^{-2}$ and $\sim$0.3 of
the Eddington limit.  For the bursting soft X-ray
transients (\cite {Whi84}; see \cite {Cam98} for recent review) the burst
activity is usually detected during the occurrence of outburst episodes,
which show peak luminosity of up to $\sim$10$^{38}$~erg~s$^{-1}$ and 
often recur on time scales of $\sim$~1 to $\sim$~10 years. 

SAX~J1750.8-2900 shows this type of transient phenomenology.
The outburst light curve has a rather clear fast rise and exponential decay
shape. In spite of the incomplete sampling it evidently shows 
variable decay behaviour, as observed in other LMXB transients (\cite
{Che97}). So there is no real evidence that the source 
had a secondary outburst after MJD~50535 as it could appear at a first glance.
The X-ray flux decreased of a factor $\approx$~20 in a period of
$\sim$~3 weeks in spring 1997, after which 
the source remained undetected (the {\em RossiXTE}~ASM and {\em BeppoSAX}
data do not show 
any other evident outburst in the period 1996 to 1999 May). We provide 
evidence that SAX~J1750.8-2900 has been observed bursting only whenever  the
intensity was above $\sim$~10 mCrab, and that the burst frequency was
positively correlated with the persistent emission (at least when the
persistent flux was in the range $\sim$~10 to $\sim$~50 mCrab). The spectral
softening seen by analysis of burst profiles is an evidence  that
SAX~J1750.8-2900 is a type-I burster, and hence that the compact object is a
neutron star. The observed peak luminosity of bursts suggests an upper limit
of 7 kpc on the source distance.  Due to the lack of optical identification
and/or  visible X-ray modulation it is not possible to classify with certainty
the binary source as a low mass system. Nevertheless, the detection of type-I
bursts is sufficient to firmly set SAX~J1750.8-2900 as a candidate 
member of the LMXB class. 

The current sample of known 
LMXB could be biased towards bright X-ray transients, due to
instrument selection effects and the established occurrence of weak, 
short lasting transients with 
long recurrence time (like e.g., 2S~1803-245, \cite {Mul98};
and SAX~J1748.9-2021, \cite {Int99}). In
fact, the recent observations by {\em BeppoSAX} and {\em RXTE} are
significantly growing the number of weak LMXB. Among them, most are NS  
transients which are also burst sources and often show high energy tails.
For this reason, these have been suggested as a possible new
subclass of low mass binaries (\cite {Hei99}). 
Indeed these systems could be  
NS soft X-ray transients (of the type of Cen~X-4 or Aql~X-1), which are 
harboured within the Galactic Bulge at quite large distances. This is what 
should be expected, as increasing the sensitivity and coverage 
will push the limit of observable distances up to  a range in
which many more sources are available, due to their  
concentration towards the Galactic Centre.


\acknowledgments

We thank the staff of the {\em BeppoSAX Science Operation Centre} and {\em Science
Data Centre} for their help in carrying out and processing the WFC Galactic Centre
observations. The {\em BeppoSAX} satellite is a joint Italian and Dutch program.
A.B., M.C., L.N. and P.U. thank Agenzia Spaziale Nazionale ({\em ASI}) for grant 
support.

\clearpage




\clearpage
 
\begin{deluxetable}{crrrrrrrrrrr}
\footnotesize
\tablecaption{Summary of spectral fitting for persistent emission}
\label{tbl-1} \tablewidth{0pt}
\tablehead{
         \colhead{Period MJD} 
       & \colhead{Model}
       & \colhead{Model Parameter}
       & \colhead{Flux 2-30 keV\tablenotemark{a}}
       & \colhead{$\chi^2_\nu$\tablenotemark{b} } 
}   
\startdata
50520.52-50521.68 & Power Law & $\Gamma$ = 2.40$\pm$0.17 & 
15.6$\pm$0.9 & 0.6\\ 
             & Bremsstrahlung & {\em k}T (keV) = 9.2$\pm$1.1 &  & 0.5 \\ 
50521.68-50522.35 & Power Law & $\Gamma$ = 2.42$\pm$0.25 &
14.1$\pm$1.2 & 1.0\\ 
             & Bremsstrahlung & {\em k}T (keV) = 9.1$\pm$1.6 &  & 1.0 \\
50525.04-50526.19 & Power Law & $\Gamma$ = 2.70$\pm$0.08 & 
8.46$\pm$0.4  & 1.5\\ 
             & Bremsstrahlung & {\em k}T (keV) = 6.1$\pm$0.3 &  & 1.0 \\ 
50526.19-50527.35 & Power Law & $\Gamma$ = 2.76$\pm$0.14 & 
5.12$\pm$0.3 & 1.1\\ 
             & Bremsstrahlung & {\em k}T (keV) = 4.3$\pm$0.4 &  & 0.9 \\ 
50527.35-50527.77 & Power Law & $\Gamma$ = 2.78$\pm$0.23 & 
4.69$\pm$0.2  & 1.0\\ 
             & Bremsstrahlung & {\em k}T (keV) = 6.0$\pm$0.7 &  & 1.0\\ 
50530.21-50531.04 & Power Law & $\Gamma$ = 3.59$\pm$0.50 & 
1.79$\pm$0.2  & 0.6\\ 
             & Bremsstrahlung & {\em k}T (keV) = 3.4$\pm$0.5 &  & 0.5 \\ 

\enddata
\tablenotetext{a}{units of 10$^{-10}$erg~cm$^{-2}$~s$^{-1}$}   
\tablenotetext{b}{reduced $\chi^2$ for 24 d.o.f.}

\end{deluxetable}




\begin{deluxetable}{crrrrrrrrrrr}
\footnotesize
\tablecaption{Parameter fit results of bursts}
\label{tbl-2} \tablewidth{0pt}
\tablehead{
         \colhead{Burst id.} 
       & \colhead{Time\tablenotemark{a}}
       & \colhead{Peak flux\tablenotemark{b}}
       & \colhead{$KT_{\rm col}$(keV)}
       & \colhead{$R_{\rm bb}$\tablenotemark{c}}    
       & \colhead{$\chi^2_{\nu}$\tablenotemark{d}}
       & \colhead{$\tau_l$\tablenotemark{e}}
       & \colhead{$\tau_h$\tablenotemark{e}}
} 
\startdata
A \tablenotemark{f} &  0.48151 & 1.2 $\pm$ 0.2 &1.6 $\pm$ 0.3 &10.4$\pm$3.2
&1.3 & 10.1 $\pm$ 5.6& 2.5 $\pm$ 1.4\\ 
B \tablenotemark{f} &  1.25676 & 3.9 $\pm$ 0.5 &2.3 $\pm$
0.2 & 8.3$\pm$1.2 &0.8 &  7.5 $\pm$ 2.6& 2.4 $\pm$ 0.6\\ 
C &  1.42773 & 3.8 $\pm$ 0.6 &2.4 $\pm$ 0.2 & 8.0$\pm$1.1 &0.9 &  5.0 $\pm$
1.6& 3.0 $\pm$ 1.0\\ D &  1.62039 & 2.2 $\pm$ 0.3 &2.2 $\pm$ 0.2 &10.7$\pm$2.1
&1.2 &  5.2 $\pm$ 2.4& 2.5 $\pm$ 0.7\\ E &  1.82726 & 4.5 $\pm$ 0.7 &2.1 $\pm$
0.2 & 9.2$\pm$1.4 &0.5 &  4.1 $\pm$ 1.7& 1.4 $\pm$ 0.4\\ F &  2.38799 & 5.4
$\pm$ 0.8 &2.2 $\pm$ 0.1 & 9.3$\pm$1.2 &1.5 &  4.5 $\pm$ 1.2& 1.5 $\pm$ 0.4\\ 
G &  2.63415 & 2.7 $\pm$ 0.4 &2.4 $\pm$ 0.2 & 7.4$\pm$1.1 &1.3 & 11.2 $\pm$
6.1& 4.4 $\pm$ 1.0\\ H & 12.41515 & 4.7 $\pm$ 0.6 &2.4 $\pm$ 0.2 & 8.4$\pm$1.2
&1.0 &  5.4 $\pm$ 2.1& 2.2 $\pm$ 0.6\\ I & 13.30355 & 4.2 $\pm$ 0.5 &2.4 $\pm$
0.2 & 7.8$\pm$1.2 &1.0 &  6.7 $\pm$ 2.1& 3.1 $\pm$ 0.9\\

\enddata
\tablenotetext{a}{time of burst rise in days (MJD-50525)} 
\tablenotetext{b}{in units of $10^{-8}$ erg~cm$^{-2}$~s$^{-1}$~, 2-26 keV} 
\tablenotetext{c}{for source at 10 kpc distance}
\tablenotetext{d}{reduced $\chi^2$ for 23 d.o.f.}
\tablenotetext{e}{$\tau_l$ is the decay time in the 2-8 keV energy band, and $\tau_h$ the decay time 
 in the 8-26 keV energy band.}
\tablenotetext{f}{burst detection affected by the earth atmosphere}

\end{deluxetable}


\clearpage

\begin{figure}
\plotone{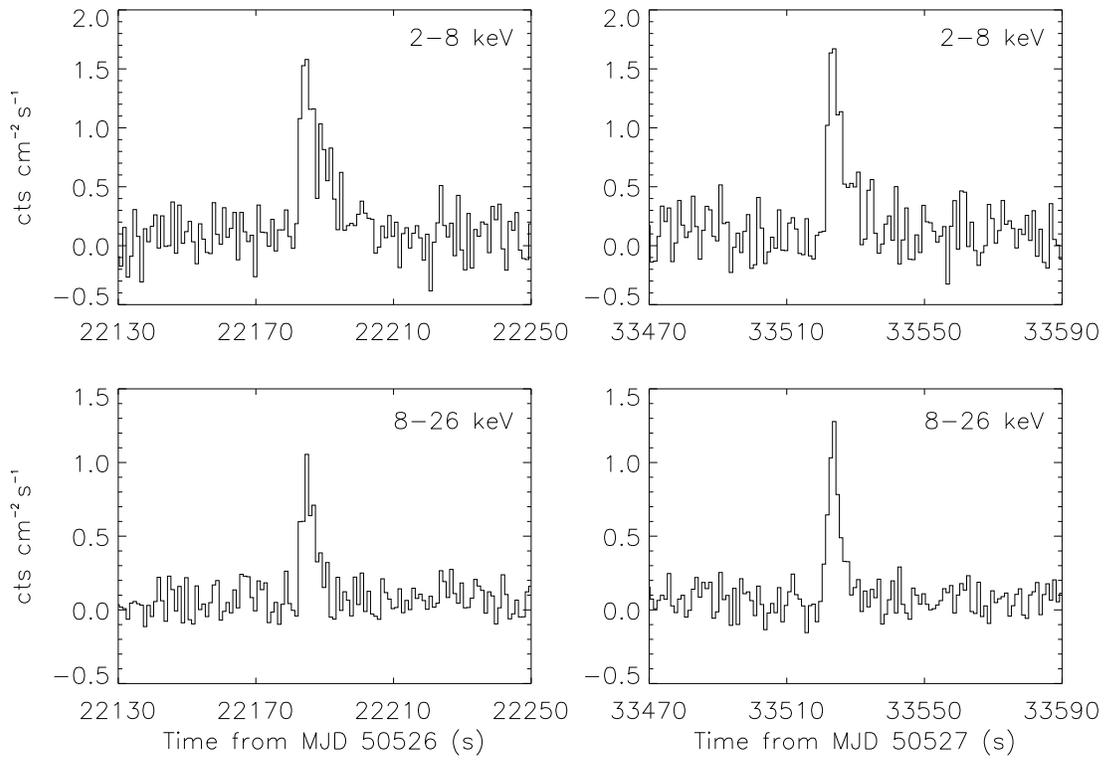}
\caption{
     Time profiles of two X-ray bursts from SAX~J1750.8-2900 detected 
     on 1997 March~19th and March~20th (from left to right).
      \label{fig3}}
\end{figure}

\end{document}